\documentclass[12pt]{article}
\usepackage{times}
\usepackage{graphics} 
\def\refn{\noindent\hangafter=1\hangindent=1cm}

\input epsf
\begin{document}
\renewcommand{\thefootnote}{\fnsymbol{footnote}}
\setcounter{footnote}{1}
\centerline{\large\bf Pulsating Components of Eclipsing Binaries}
\centerline{\large\bf in the ASAS-3 Catalog}

\vspace{0.3cm}
\centerline{by}

\vspace{0.3cm}
\centerline{A.~P i g u l s k i\quad and\quad G.~M i c h a l s k a}

\vspace{0.3cm}
\centerline{Astronomical Institute, University of Wroc{\l}aw}
\centerline{Kopernika 11, 51-622 Wroc{\l }aw, Poland}
\centerline{E-mail: (pigulski,michalska)@astro.uni.wroc.pl}

\vspace{0.5cm}
\centerline{\it Received }

\vspace{1cm}
\centerline{ABSTRACT}

\vspace{0.5cm}
{\small
As a result of the search among about 11\,000 stars from the public ASAS-3 database, we report detection of pulsating
components in eleven eclipsing binaries.  In particular, we have found three classical Algols, MX~Pav, IZ~Tel, and VY~Mic, 
with $\delta$~Sct-type primary components. In six other eclipsing binaries, the short-period variability can also be 
interpreted in terms of $\delta$~Sct-type pulsations, but in these systems both components are probably main-sequence 
stars.  In HD\,99612, the pulsation mode shows significant amplitude decrease during the time interval covered by observations.  
In addition, we find variability in one of the components of the eclipsing and double-lined spectroscopic O-type binary ALS\,1135 
which we interpret as a $\beta$~Cep-type pulsation. Finally, we find Y Cir to be a good candidate for an SPB star in an eclipsing 
binary system.

\vspace{0.2cm} 

}

\vspace{0.5cm}
\centerline{\bf 1.~Introduction}

\vspace{0.5cm}
It is well known that a combination of the light curve of an eclipsing binary with its double-lined spectroscopic orbit allows
to derive masses and radii of the components. These parameters are crucial in many astrophysical applications including
the study of pulsating stars by means of asteroseismology because it requires a good knowledge of stellar parameters of the
stars to be studied. For this reason, eclipsing binaries with pulsating components are favored in the asteroseismic work.

Almost all types of pulsating stars are found as members of eclipsing binary systems. An overview of known pulsating
components in eclipsing systems was recently given by Pigulski (2005a), while several open questions in this topic were
discussed by Lampens (2005). Since $\delta$~Sct stars are rather common, they form the largest group of
pulsating stars in eclipsing binaries. A catalogue of 25 eclipsing systems with $\delta$~Sct components was recently
published by Soydugan {\it et al}.~(2006, hereafter So06). These authors also presented a list of eclipsing binaries with 
at least one of the components located in the $\delta$~Sct instability strip.

Most of $\delta$ Sct stars in the So06's list are the primaries of semi-detached Algol-type eclipsing systems in which 
mass transfer can be still in work. These main-sequence mass-accreting pulsating components are called ‘oscillating EA’ (oEA) stars 
(Mkrtichian {\it et al}.~2002, 2004). The internal structure and pulsating properties of these stars might be different from those of a
single star. It would be therefore interesting to compare pulsating properties of oEA stars and single $\delta$~Sct stars. 
This is why oEA stars are now intensively searched for (Kim {\it et al}.~2003, 2005; Mkrtichian {\it et al}.~2005, 2006).

There are at least two other reasons justifying searches for pulsating components of eclipsing binaries. First, there is a 
possibility of using eclipses for mapping non-radial mode(s) in a pulsating star (e.g., B\'{\i}r\'o and Nuspl 2005), especially 
because in eclipsing systems we know the inclination and it may be assumed that the rotation axis is perpendicular to the orbital 
plane. The method works best for sectoral modes and has been applied to several pulsating stars (Rodr\'{\i}guez 
{\it et al.}~2004a,b; Reed {\it et al}.~2005). For a successful mode identification, however, a large number of eclipses has 
to be observed.

Finally, the effects of tidal interaction on pulsating modes can be studied in an eclipsing system with a pulsating component.
There are stars in which frequencies close to integer multiples of the orbital period were found (e.g., Handler {\it et al}.~2002) 
and which were interpreted as tidally excited. In many Algols, the rotational period of the primary is not synchronized 
with the orbital period. A spectroscopically determined rotational velocity of these components may be overestimated because 
of a rapidly rotating accreting matter. As discussed by Mkrtichian {\it et al}.~(2002), an investigation of the rotational 
splitting of non-radial pulsation modes in oEA systems can provide an accurate measurement of the internal rotation, as well 
as an estimate of surface differential rotation.

Since eclipsing systems with pulsating components are excellent targets for asteroseismic studies, we searched the All Sky 
Automated Survey (ASAS) database in the hope of finding new such objects among many eclipsing binaries discovered in 
this excellent survey.

\vspace{0.8cm}
\centerline{\bf 2.~The Data and Analysis}

\vspace{0.5cm}
We used data from the public database of the third part of the All Sky Automated Survey, ASAS-3 (Pojma\'nski 2001, 2002,
2003; Pojma\'nski and Maciejewski 2004, 2005; Pojma\'nski {\it et al}.~2005). These observations were carried out in the years
2000--2006 at the Las Campanas Observatory. The database contains $V$-filter light curves of over 50,000 variable stars 
with declinations below +28$^{\circ}$.

A preliminary classification of variability of stars in the ASAS catalogue was made by its authors in an automatic way.
Using Fourier coefficients, they divided eclipsing binaries into three classes: detached (ED), semi-detached (ESD) and contact (EC).
Some stars were given a double or multiple classification.  In total, over 11\,000 variable stars were classified as eclipsing binaries.
This sample was subject of our detailed analysis.

As to the choice of aperture, combination of subsets and removing outliers, the current analysis followed the procedure described
in detail by Pigulski (2005b). Next, however, the data were phased with the orbital period calculated by means of the AoV program
(Schwarzenberg-Czerny 1996) in a narrow range of frequencies, using the orbital frequency from the public ASAS database
as a starting point. The light curve phased with the orbital period was then fitted by a smooth function, which was a cubic spline 
calculated using averages in phase intervals equal to 0.05 or 0.025.  This smooth curve was subtracted from the original data.
The residuals obtained in this way were then searched for periodic signals non-commensurable with the orbital frequency or its aliases.  
This was done by calculating Fourier periodograms in the range between 0 to 80~d$^{-1}$. If the signal-to-noise ratio for the highest 
peak in the periodogram exceeded five, the light curve was examined in detail. In total, we found twelve stars which might be 
regarded as eclipsing binaries with pulsating components. All but FR Sct are described below. The results for FR~Sct have been 
published separately (Pigulski and Michalska 2007) because this system is a unique triple VV~Cep star and its quasi-periodic 
variability has a time scale much longer than that of stars described below. Moreover, it is not certain that these variations are due 
to pulsations.

\begin{center}
{\small
T a b l e\quad 1\\
Eclipsing binaries with pulsating components found in the ASAS-3 catalogue
}

\vspace{2mm}
\footnotesize
\begin{tabular}{crccc}
\hline\noalign{\smallskip}
ASAS  &  \multicolumn{1}{c}{$V$}& MK spectral& Other & Type of  pulsating \\
designation&  \multicolumn{1}{c}{[mag]} &classification &designation& component \\
\noalign{\smallskip}\hline\noalign{\smallskip} 
182411--6356.9 & 11.35 & --- & MX Pav    & $\delta$ Sct, oEA \\
202844--5620.8 & 12.06 & --- & IZ Tel    & $\delta$ Sct, oEA \\
204907--3343.9 &  9.47 & A4 III/IV & VY Mic    & $\delta$ Sct, oEA \\	
\noalign{\smallskip}\hline\noalign{\smallskip}	
073904--6037.2 & 10.03 & A5 II & CPD $-$60$^\circ$871  & $\delta$ Sct \\
074422--0641.8 &  8.73 & F0 V    & HD 62571  & $\delta$ Sct \\
110615--4224.6 & 10.47 & ---   & CPD $-$41$^\circ$5106 & $\delta$ Sct/$\beta$ Cep \\
112733--2450.2 & 10.98 & A7 II/III:  & HD 99612  & $\delta$ Sct \\
232548--1136.6 &  9.55 & A2 III       & HD 220687 & $\delta$ Sct \\
234520--3100.5 & 10.95 & ---   & CPD $-$31$^\circ$6830 & $\delta$ Sct \\
\noalign{\smallskip}\hline\noalign{\smallskip}
084350--4607.2 & 10.56 & O6.5\,V((f))+B1 V & ALS 1135 & $\beta$ Cep: \\
\noalign{\smallskip}\hline\noalign{\smallskip}
133910--6502.2 & 11.21 & --- & Y Cir   & SPB: \\
\noalign{\smallskip}\hline
\end{tabular}
\end{center}

\vspace{0.8cm}
\centerline{\bf 3.~The Results}

\vspace{0.5cm}
As stated above, we found eleven stars which are very good candidates for pulsating components of eclipsing binaries. They are 
listed in Table 1. In two cases (HD\,62571 and 220687), more than one periodicity was found; this is an important argument in favor 
of pulsations as an explanation of the short-period variability. For the remaining stars, we found only a single periodicity. Their
periods are unrelated to the orbital periods and are very short in most cases. They are, therefore, very likely to be due to pulsations.
We assign the type of variability (given in the last column of Table 1) on the basis of the values of detected frequencies and spectral
type(s) of the component(s). In most cases, the available MK spectral type is that of the primary. In systems with very large difference in eclipse
depths there is little doubt that it is the primary (brighter) component which pulsates. Otherwise the reduction of the amplitude in the
presence of a brighter non-pulsating primary would lead to a significant reduction of the amplitude of the pulsating secondary and
possibly prevent detection. This argument is strongest in Algols, where the secondary is an evolved cool star, not expected to show
short-period pulsations.

Taking into account frequencies of the short-period variability which we attribute to pulsations,
available spectral types, multicolor photometry, the shape of the light curve and eclipse depths,
we can divide the eleven systems listed in Table 1 into four distinct groups. Tables 2 and 3 summarize
the properties of the eclipsing light-curves and the pulsation(s), respectively. In Table 2 and 3, $T_0$ = HJD\,2450000.0.
In Table 2, $T_{\rm min}$~I stands for the time of the primary minimum. In Table 3, $N_{\rm obs}$ denotes the number
of observations, $A$, $V$-filter semi-amplitude, $T_{\rm max}$, time of maximum light, $\sigma_{\rm res}$, standard
deviations of residuals, and DT, detection threshold defined as signal-to-noise equal to four.

\begin{center}
{\small
T a b l e\quad 2\\
Parameters of the eclipses for eleven stars of Table 1
}

\vspace{2mm}
\footnotesize
\begin{tabular}{ccccc}
\hline\noalign{\smallskip}
&Orbital &$T_{\rm min}$ I $- T_0$& \multicolumn{2}{c}{Eclipse depth [mag]} \\
Star & period [d] & [d]& primary & secondary \\
\noalign{\smallskip}\hline\noalign{\smallskip} 
MX Pav & 5.730835(04) & 2074.5266(07) & 1.96 & 0.13 \\
IZ Tel & 4.880219(04) & 2086.8791(08) & 1.99 & 0.12\\
VY Mic & 4.436373(02) & 2079.0409(05) & 0.99 & 0.08\\
\noalign{\smallskip}\hline\noalign{\smallskip} 
CPD\,$-$60$^\circ$871 & 1.2209619(7) & 1978.6871(07) & 0.14 & 0.11\\
HD\,62571 & 3.208647(10) & 1950.6576(27) & 0.12 & 0.04\\
CPD\,$-$41$^\circ$5106 & 2.136998(04) & 1950.2300(20) & 0.18 & 0.10 \\
HD\,99612 & 2.778758(04) & 1965.9735(08) & 0.46 & 0.07\\
HD\,220687 & 1.594251(03) & 2058.5821(13) & 0.21 & 0.03\\
CPD\,$-$31$^\circ$6830 & 0.8834334(9) & 2069.1521(10) & 0.22 & 0.13 \\
\noalign{\smallskip}\hline\noalign{\smallskip} 
ALS\,1135 & 2.753173(10) &1951.7719(38) & 0.15 & 0.11 \\
\noalign{\smallskip}\hline\noalign{\smallskip}
Y Cir & 3.169971(03) & 2167.6588(06) & 1.12 & 0.08\\
\noalign{\smallskip}\hline
\end{tabular}
\end{center}

\begin{center}
{\small
T a b l e\quad 3\\
Parameters of short-period variability for eleven stars of Table 1
}

\vspace{2mm}
\footnotesize
\begin{tabular}{cccccccc}
\hline\noalign{\smallskip}
 & &\multicolumn{1}{c}{Frequency} & \multicolumn{1}{c}{$A$} & \multicolumn{1}{c}{$T_{\rm max} - T_0$} & 
\multicolumn{1}{c}{$\sigma_{\rm res}$} & \multicolumn{1}{c}{DT}\\
Star name &  $N_{\rm obs}$ & \multicolumn{1}{c}{[d$^{-1}$]} & \multicolumn{1}{c}{[mmag]} & \multicolumn{1}{c}{[d]}
& \multicolumn{1}{c}{[mmag]} & \multicolumn{1}{c}{[mmag]} \\
\hline\noalign{\smallskip} 
MX Pav &  652 & 13.227222(10) & 76.9(28) & 2877.6902(04) & 49.4 & 14.1\\
IZ Tel & 551 & 13.558007(23) & 45.9(38) &  2831.1936(10) & 60.2 & 18.1\\
VY Mic & 274 & 12.234111(29) & 19.4(20) & 2815.1716(13) & 22.3 & 8.9\\
\noalign{\smallskip}\hline\noalign{\smallskip} 
CPD\,$-$60$^\circ$871 & 498 & 4.667898(13) & 17.1(11) & 2994.1667(21) & 16.6 & 5.3\\
HD 62571 & 464 & 9.051421(08) & 41.7(12) & 2834.9212(05) & 17.7 & 5.8\\
&  & 8.428109(37) & 9.1(12) & 2834.9300(24) & &\\
CPD\,$-$41$^\circ$5106 & 326 & 8.247307(23) & 20.2(17) & 2859.8691(17) & 21.6 & 8.5\\
HD 99612 & 587 & 14.713698(23) & 8--35, Var & 2802.4009(10) & 24.5 & 7.1\\
HD\,220687 & 329 & 26.169254(35) & 12.8(14) & 2810.6042(07) & 16.7 & 6.8\\
CPD\,$-$31$^\circ$6830 & 273 & 5.463348(12) & 54.1(21) & 2731.6486(11) & 23.6 & 10.4 \\
\noalign{\smallskip}\hline\noalign{\smallskip} 
ALS\,1135 & 346 & 2.310923(24) & 23.8(22) & 2851.2250(62) & 27.7 & 10.5\\
\noalign{\smallskip}\hline\noalign{\smallskip}
Y Cir & 378 & 0.901804(32) & 29.6(27) & 2859.263(17) & 35.3 & 12.8\\
\hline
\end{tabular}

\end{center}

\begin{figure}[!ht]
\centering
\epsfxsize=5.5in
\epsfbox{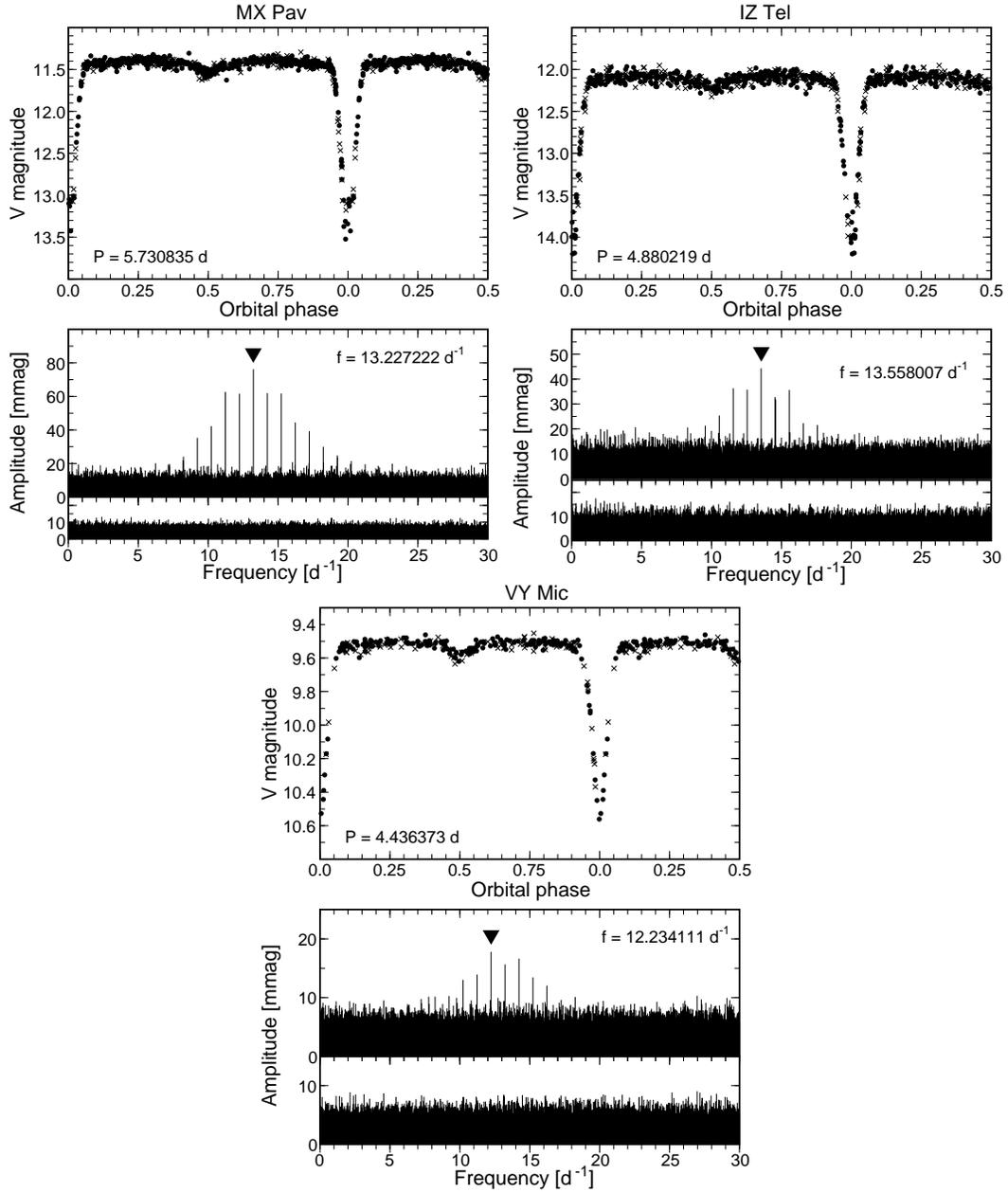}
\caption{\small The light curves of three oEA systems included in Group 1: MX~Pav, IZ~Tel and VY~Mic. For each star, the upper panels show
the eclipsing light-curve phased with period and initial epoch given in Tab.~2. Data of higher accuracy are plotted with dots, those of lower
accuracy, with crosses.  The lower panels show Fourier spectra for consecutive steps of prewhitening. The first spectrum is the spectrum 
calculated for residuals from the eclipse light curve. Detected frequencies, given in Tab.~3, are indicated with inverted triangles.}
\end{figure}

\vspace{0.5cm}
{\it 3.1.~Group 1: Oscillating EA Stars}

\vspace{0.3cm} 
The first group consists of three stars, MX~Pav, IZ~Tel, and VY~Mic, which are obvious oEA systems. As can be seen from Fig.~1
(see also Table 1), all three binaries have similar light curves with very deep primary and shallow secondary eclipses.  In all these
cases, after freeing the data from eclipses, the Fourier periodograms revealed a single periodicity with a frequency typical for 
$\delta$~Sct stars (see Table 3). Of the three stars, only VY~Mic has its MK spectral classification available, A4\,III/IV (Houk 1982). 
For the other two stars, Svechnikov and Kuznetsova (1990) provided spectral types (A5)+[K3\,IV] and (A8)+[G8\,IV], respectively.
These are not spectroscopically determined types, however. For primaries, they were estimated from the statistical dependences
for eclipsing binaries of different types, for secondaries, computed from the observed values of the surface brightness ratio. 
Nevertheless, for all three stars the spectral types are consistent with the assumption that the primary is a $\delta$~Sct star.
It is interesting to note that VY~Mic was listed by So06 in their list of candidate semi-detached oEA systems.

\begin{figure}[!t]
\centering
\epsfxsize=5.5in
\epsfbox{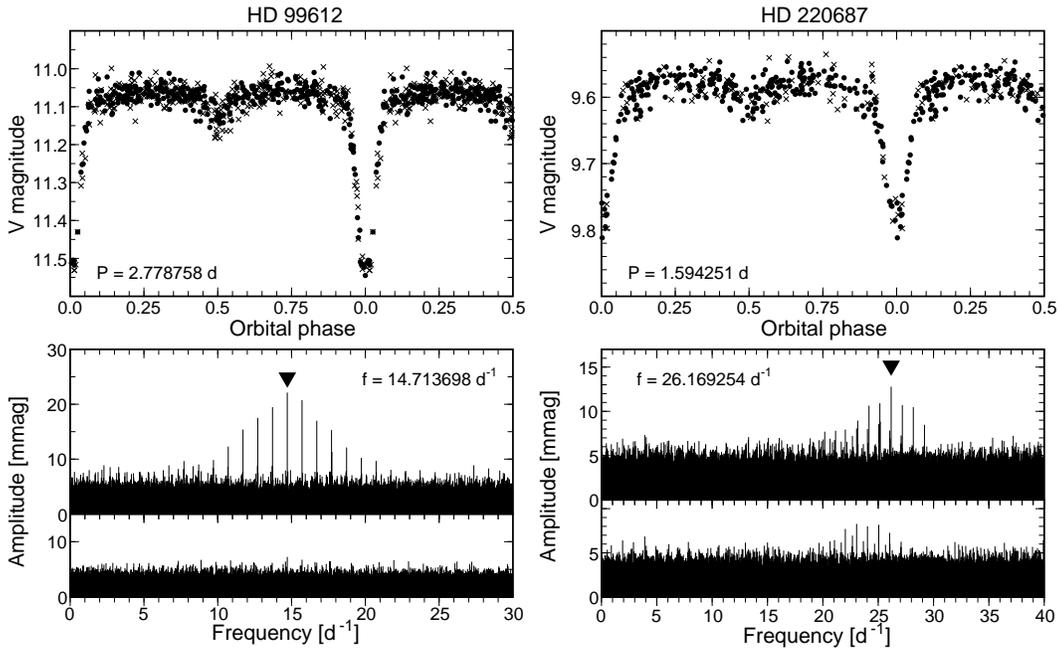}
\caption{\small The same as in Fig.~1, but for two systems included in Group 2: HD\,99612 and HD\,220687.} 
\end{figure}
\begin{figure}[!t]
\centering
\epsfxsize=5.5in
\epsfbox{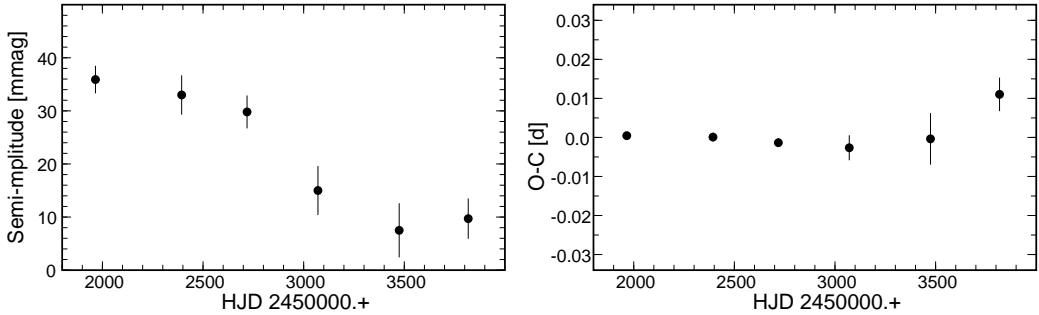}
\caption{\small Left panel: the amplitude change of the pulsating mode in HD\,99612. Right panel: the O$-$C diagram for the times of maximum light in six
subsamples. The ordinate range is equal to the pulsation period.} 
\end{figure}

It would be worthwhile to check if the {\it known} oEA systems can be confirmed using the ASAS data. Ten such systems
have declinations below +28$^{\circ}$, i.e., were covered by the ASAS observations. Five of them, TZ~Eri, WX~Eri, AS~Eri, RX~Hya, 
and AO~Ser, are included in the ASAS-3 catalogue. We analyzed their ASAS photometry, but failed to find short-period variability.
For all these stars but WX~Eri this was
simply because amplitudes of pulsations reported in the literature were smaller than the detection threshold of the ASAS-3 data.
For WX~Eri, which was recently renamed V1241~Tau (Kazarovets {\it et al}.~2006), Sarma and Abhyankar (1979)
claimed $\delta$~Sct-type pulsations with periods five and six times shorter than the orbital period. However, they performed
rectification fitting the light curve of the star with a Fourier series truncated to the first three components. We conclude therefore
that the short-period term in V1241~Tau they found is just a harmonic of the orbital frequency which appeared as an artefact of
the rectification they done.  Arentoft {\it et al}.~(2004) did not find any evidence for pulsations in this system. We reached the same
conclusion from the analysis of the ASAS data: no significant signal was found in the residuals. Therefore, the star should be no longer
regarded as an oEA system.

\vspace{0.5cm}
{\it 3.2.~Group 2: Detached Systems with $\delta$~Sct Components}

\vspace{0.3cm} 
There are six stars in the second group. In general, they show much shallower primary eclipses than the oEA systems. As can be
seen in Fig.~2, in two, HD\,99612 and 220687, the difference in the depth between the primary and secondary eclipse is still quite large 
On the other hand, however, it seems that the primary minima for both stars have a flat bottom. Consequently, we may argue that
the primary eclipse is most likely a transit, and the secondary component is a main-sequence star, smaller and thus less massive
than the primary.  HD\,99612 is classified as A7\,II/III:, HD\,220687 as A2\,III (Houk and Smith-Moore 1988); thus pulsations, 
presumably of the primary components, can be attributed to $\delta$~Sct-type variability. The Fourier spectrum of HD\,220687
shown in Fig.~2 clearly shows at least one more periodicity. Its frequency equals to 23.0413~d$^{-1}$ or one of its daily aliases.
Since the value of the true frequency is uncertain, we did not include the secondary frequency in the solution provided in Tab.~3.

Surprisingly, we found the amplitude of the pulsation in HD\,99612 to decrease with time (Fig.~3). Since there is no clear evidence for
an associated period change (see the right panel of Fig.~3), this must be a real change of amplitude, not an apparent change due
to the modulation of modes with very close frequencies. The amplitude changes are not uncommon among $\delta$~Sct stars (e.g., 
Breger 2000).

\begin{figure}[!t]
\centering
\epsfxsize=5.5in
\epsfbox{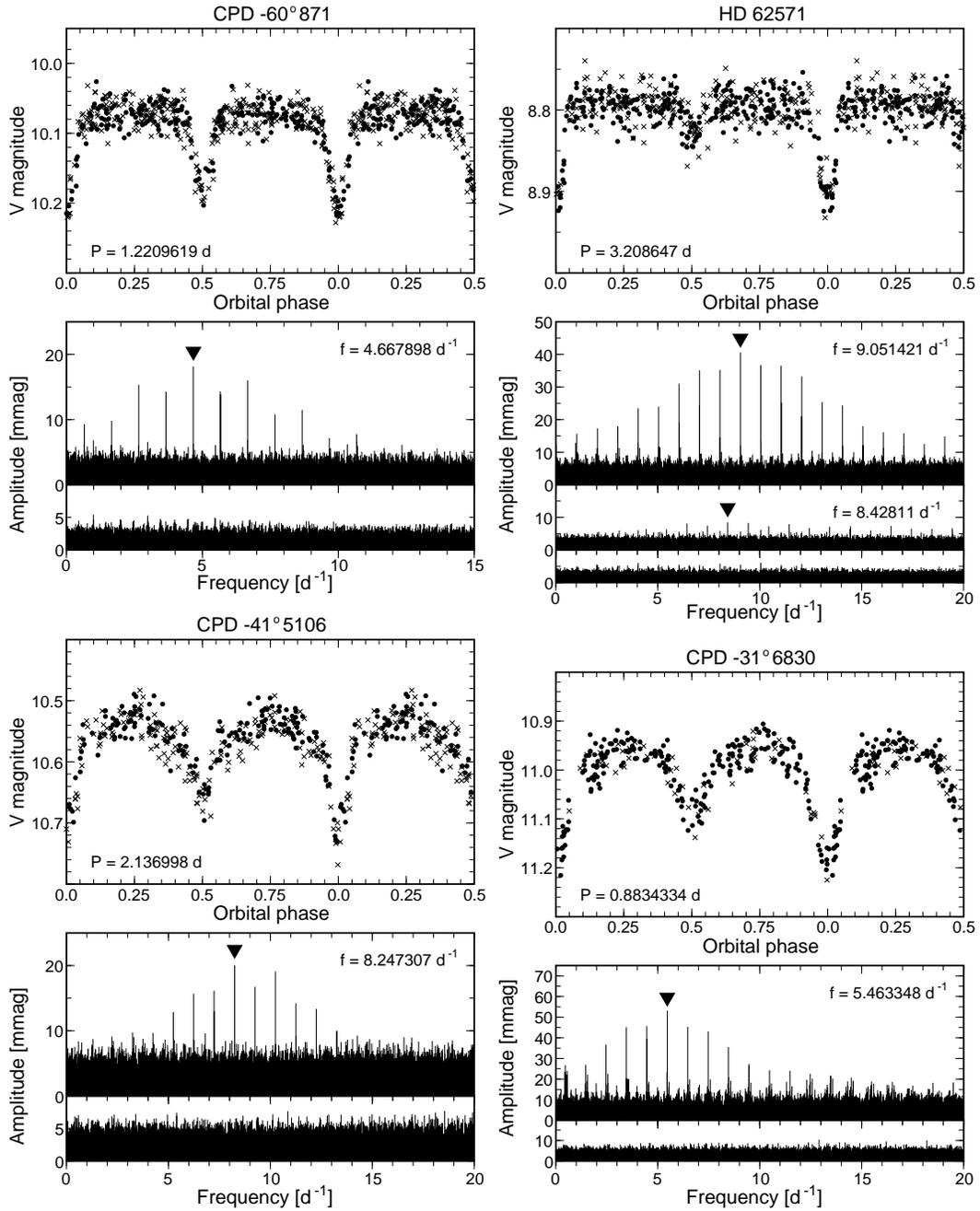}
\caption{\small The same as in Fig.~1, but for the four systems, CPD\,$-$60$^\circ$871, HD\,62571, CPD\,$-$41$^\circ$5106, and
CPD\,$-$31$^\circ$6830, included in the second group.}
\end{figure}

In the remaining four stars of this group, shown in Fig.~4, the primary and secondary eclipses have comparable depths. Probably all are detached
 systems, although in two, CPD\,$-$41$^\circ$5106 and CPD\,$-$31$^\circ$6830, the reflection effect is also visible. For these four stars it is
difficult to conclude which component pulsates. As far as the type of variability is concerned, probably all are $\delta$~Sct stars. For two
(see Tab.~1) this in agreement with their MK classification. For the remaining two, CPD\,$-$41$^\circ$5106 and CPD\,$-$31$^\circ$6830,
spectral types are not available, so that the alternative would be $\beta$~Cep-type of variability. However, both are located far from
the Galactic plane, 17$^\circ$ and 75$^\circ$, respectively. Thus, at least for CPD\,$-$31$^\circ$6830 it is unlikely that the primary is an early-B type
star. As can be seen in Fig.~4, in one star of this group, HD\,62571, we detected two periodic terms.

\vspace{0.5cm}
{\it 3.3.~ALS\,1135}

\vspace{0.3cm}
ALS\,1135 = CPD $-$45$^\circ$2920 is a very interesting star and a member of the OB association Bochum 7 = Vela OB3
(Moffat and Vogt 1975, Sung {\it et al}.~1999). The age of the association is estimated to be 6~Myr (Sung {\it et al}.~1999).
The association is located at a distance of about 5~kpc (Moffat and Vogt 1975, Sung {\it et al}.~1999, Corti {\it et al}.~2003).
The MK type of ALS\,1135 was first derived by Vijapurkar and Drilling (1993) as O6\,III. Recently, Corti {\it et al}.~(2003)
classified the star as of O6.5\,V((f)) type. The same authors (Corti {\it et al}.~1999, 2003) found ALS\,1135 to be a single-lined
spectroscopic binary with a period of 2.75320~d. Independently, the star was found to be eclipsing (Pojma\'nski 2003).
Following this discovery, Fern\'andez Laj\'us and Niemela (2006) re-analyzed their spectra and derived the secondary's radial
velocities close to quadratures. As a result, they found masses and radii of both components from the combined analysis
of their SB2 orbit and the ASAS light curve. The primary and secondary masses are equal to 30 $\pm$ 1 and 9 $\pm$ 1~M$_\odot$,
respectively, and the radii, 12 $\pm$ 1 and 5 $\pm$ 1~R$_\odot$. The secondary was found to be a B1\,V-type star.

\begin{figure}[!t]
\centering
\epsfxsize=5.5in
\epsfbox{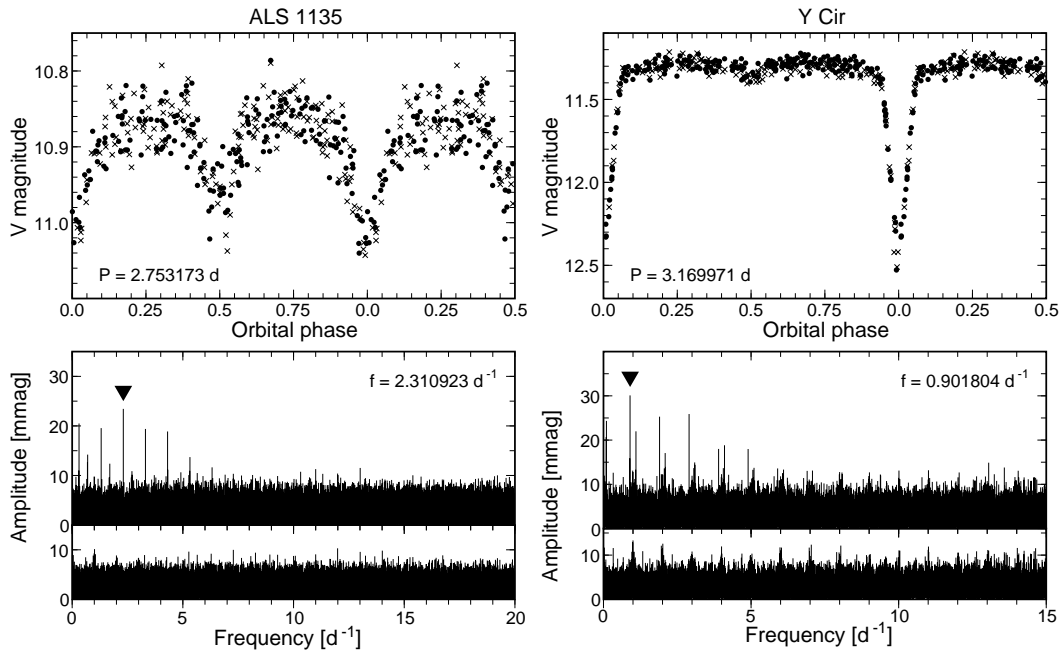}
\caption{\small The same as in Fig.~1, but for the two systems, ALS\,1135 and Y~Cir.} 
\end{figure}

We find another interesting feature in ALS\,1135. Despite the eclipses (Fig.~5), the star shows clearly residual sinusoidal variation with
a frequency of 2.3109~d$^{-1}$ (the period equal to 0.4327~d).  The period seems to be too long for a pulsation in an acoustic mode in
a 9~M$_\odot$ star (secondary component), but is acceptable if the pulsations originate in the O-type primary.
In such a massive star, the theory allows $\beta$ Cep-type pulsation with period we derived (see Pamyatnykh 1999).

\vspace{0.5cm}
{\it 3.4.~Y Circini}

\vspace{0.3cm}
The last eclipsing binary with a pulsating component we present here is Y~Cir. It shows an Algol-type light curve (Fig.~5) similar to oEA systems
included in Group 1, but the frequency detected in residuals is equal to 0.902~d$^{-1}$. Unfortunately, the MK type is not available for this star.
The only existing estimation of spectral types of the components, (A2)+[K0\,IV], comes from Svechnikov and Kuznetsova (1990)
and was done as explained in Sect.~3.1. Thus, assuming that the primary can be of late B type, the most plausible explanation of the
detected periodicity is a pulsation in a g mode typical for slowly pulsating B (SPB) stars.

\vspace{0.8cm}
\centerline{\bf 4.~Conclusions}

\vspace{0.5cm}
 It cannot be excluded that for some stars discussed in this paper, periodic variability which we attribute to pulsations
does not occur in any of the binary components but in a nearby star. The probability of such contamination in the ASAS
photometry might be significant due to the low spatial resolution of the images. At least for stars located 0.3--1$^\prime$ off the target star
this can be checked by comparing results of photometry with different apertures. We performed such a check for all eleven stars
and found no clear dependence of the resulting amplitudes on the aperture. Still, a contamination by very close stars cannot be excluded.
For example, one of the targets, CPD $-$31$^\circ$6830, is known to be a visual double with two equally bright components
separated by less than 1$^{\prime\prime}$ (Rossiter 1933).

The question of which component pulsates can be easily answered once a high-quality light-curve with good temporal
resolution is obtained. This was not possible with the ASAS data as typically only one point per night was available,
a good reason for making follow-up observations. Practically all systems we found are good targets for
asteroseismic work provided that stellar parameters of the components will be derived by simultaneous
spectroscopic and photometric observations.

The three oEA systems we discovered are among the stars with the largest amplitudes of pulsations in the whole group of
oEA stars. They are therefore very attractive for mode identificaton by means of the eclipse mapping technique.
Next, the long-term monitoring of the period changes in these stars may allow determination of the mass-transfer rate.
If rotationally split high-degree modes will be discovered, the rotational period of the star could be derived. This could contribute
to explaining asynchronous rotation in Algols.

Although O-type stars were searched for $\beta$~Cep-type pulsations (Balona 1992, Pigulski and Ko{\l}aczkowski 1998), the
evidence for pulsations in these stars was provided rather by the line-profile variations than photometry (Fullerton {\it et al}.~1996).
Recently, three pulsating O-type stars were found using the ASAS photometry (Pigulski and Pojma\'nski, in preparation), $\beta$~Cep-type pulsations
were also confirmed in $\zeta$~Oph, a fast-rotating O9.5\,Ve star (Walker {\it et al}.~2005). However, all these stars are of late O type.
If the pulsations in ALS\,1135 were confirmed, the pulsating O6.5\,V-type component would be unique among $\beta$~Cep
stars.

In the prelimary paper showing results of this work (Michalska and Pigulski 2007), we claimed that three other stars,
HD\,94529, HD\,251158 and V4396\,Sgr, are eclipsing binaries with pulsating components. A more detailed
analysis in which some new data from the ASAS survey were included showed that: (i) the short-period
variability of HD\,94529 is an alias of the harmonic of the orbital frequency; (ii) the variability in two remaining stars
(with periods of the order of 1 day) is not related to the orbital period but should be interpreted rather in terms of the ellipsoidal
variation. The systems might therefore be interesting because they may be multiple, but they are not eclipsing systems
with pulsating components.

\vspace{0.5cm}
{\bf Acknowledgements.} This work was supported by the KBN grant 1\,P03D\,016\,27.  We thank Prof.~M.\,Jerzykiewicz for
fruitful comments and discussion. The research has made use of the SIMBAD database, operated at CDS, Strasbourg, France.

\vspace{0.5cm}
\centerline{REFERENCES}
\vspace{0.3cm}
{\small
\refn Arentoft, T., Lampens. P., Van Cauteren, P., Duerbeck, H.W., Garc\'{\i}a-Melendo, E., Sterken, C., 2004, {\it Astron.~Astrophys.}, {\bf 418}, 249.\par
\refn Balona, L.A., 1992, {\it M.N.R.A.S.}, {\bf 254}, 404.\par
\refn B\'ir\'o, I.B., Nuspl, J., 2005, {\it A.S.P.~Conf.~Ser.}, {\bf 333}, 221.\par
\refn Breger, M., 2000, {\it A.S.P.~Conf.~Ser.}, {\bf 203}, 421.\par
\refn Corti, M.A., Morrell, N.I., Niemela, V.S., 1999, {\it Bol.~Asoc.~Argentina de Astron.}, {\bf 43}, 32.\par
\refn Corti, M., Niemela, V., Morrell, N., 2003, {\it Astron.~Astrophys.}, {\bf 405}, 571.\par
\refn Fern\'andez Laj\'us, E., Niemela, V.S., 2006, {\it M.N.R.A.S.}, {\bf 367}, 1709.\par
\refn Fullerton, A.W., Gies, D.R., Bolton, C.T., 1996, {\it Astrophys.~J.~Suppl.}, {\bf 103}, 475.\par
\refn Handler, G., {\it et al}. 2002, {\it M.N.R.A.S.}, {\bf 333}, 262.\par
\refn Houk, N., 1982, {\it Michigan Spectral Survey, Ann Arbor, Dep.~Astron., Univ.~Michigan}, {\bf 3}.\par
\refn Houk, N., Smith-Moore, M., 1988, {\it Michigan Spectral Survey, Ann Arbor, Dep.~Astron., Univ.~Mi\-chi\-gan}, {\bf 4}.\par
\refn Kazarovets, E.V., Samus, N.N., Durlevich, O.V., Kireeva, N.N., Pastukhova, E.N., 2006, {\it I.B.V.S.}, {\bf 5721}.\par
\refn Kim, S.-L., Lee, J.W., Kwon, S.-G., Youn, J.H., Mkrtichian, D.E., Kim, C., 2003, {\it Astron.~Astrophys.}, {\bf 405}, 231.\par
\refn Kim, S.-L., Kim, S.H., Lee, D.-J., Lee, J.A., Kang, Y.B., Koo, J.-R., Mkrtichian, D., Lee, J.W., 2005, {\it A.S.P.~Conf.~Ser.}, {\bf 333}, 217.\par
\refn Lampens, P., 2005, {\it A.S.P.~Conf.~Ser.}, {\bf 349}, 153.\par
\refn Michalska, G., Pigulski, A., 2007, {\it Comm.~in Asteroseismology}, in press.\par
\refn Mkrtichian, D.E., Kusakin, A.V., Gamarova, A.Yu., Nazarenko, V., 2002, {\it A.S.P. Conf.~Ser.}, {\bf 256}, 259.\par
\refn Mkrtichian, D.E., {\it et al}. 2004, {\it Astron.~Astrophys.}, {\bf 419}, 1015.\par
\refn Mkrtichian, D.E., {\it et al}. 2005, {\it A.S.P.~Conf.~Ser.}, {\bf 333}, 197.\par
\refn Mkrtichian, D.E., {\it et al}. 2006, {\it Astrophys.~Space Sci.}, {\bf 304}, 169.\par
\refn Moffat, A.F.J., Vogt, N., 1975, {\it Astron.~Astrophys.~Suppl.} {\bf 20}, 85.\par
\refn Pamyatnykh, A.A., 1999, {\it Acta Astron.}, {\bf 49}, 119.\par
\refn Pigulski, A., 2005a, {\it A.S.P.~Conf.~Ser.}, {\bf 349}, 137.\par
\refn Pigulski, A., 2005b, {\it Acta Astron.}, {\bf 55}, 219.\par
\refn Pigulski, A., Ko{\l}aczkowski, Z., 1998, {\it M.N.R.A.S.}, {\bf 298}, 753.\par
\refn Pigulski, A., Michalska, G., 2007, {\it I.B.V.S.}, {\bf 5757}.\par
\refn Pojma\'nski, G., 2001, {\it A.S.P.~Conf.~Ser.}, {\bf 246}, 53.\par
\refn Pojma\'nski, G., 2002, {\it Acta Astron.}, {\bf 52}, 297.\par
\refn Pojma\'nski, G., 2003, {\it Acta Astron.}, {\bf 53}, 341.\par
\refn Pojma\'nski, G., Maciejewski, G., 2004, {\it Acta Astron.}, {\bf 54}, 153.\par
\refn Pojma\'nski, G., Maciejewski, G., 2005, {\it Acta Astron.}, {\bf 55}, 97.\par
\refn Pojma\'nski, G., Pilecki, B., Szczygie{\l}, D., 2005, {\it Acta Astron.}, {\bf 55}, 275.\par
\refn Reed, M.D., Brondel, B.J., Kawaler, S.D., 2005, {\it Astrophys.~J.}, {\bf 634}, 602.\par
\refn Rodr\'{\i}guez, E., {\it et al}. 2004a, {\it M.N.R.A.S.}, {\bf 347}, 1317.\par
\refn Rodr\'{\i}guez, E., Garc\'{\i}a, J.M., Gamarova, A.Yu., Costa, V., Daszy\'nska-Daszkiewicz, J., 
        L\'opez-Gonz\'alez, M.J., Mkrtichian, D.E., Rolland, A., 2004b, {\it M.N.R.A.S.}, {\bf 353}, 310.\par
\refn Rossiter, R.A., 1933, {\it Mem.~Royal Astron.~Soc.}, {\bf 65}, 28.\par
\refn Sarma, M.B.K., Abhyankar, K.D., 1979, {\it Astrophys.~Space Sci.}, {\bf 65}, 443.\par
\refn Schwarzenberg-Czerny, A., 1996, {\it Astrophys.~J.}, {\bf 460}, L107.\par
\refn Soydugan, E., Soydugan, F., Demircan, O., \.Ibano\v{g}lu, C., 2006, {\it M.N.R.A.S.}, {\bf 370}, 2013 (So06).\par
\refn Sung, H., Bessell, M.S., Park, B.-G., Kang, Y.H., 1999, {\it Journal of the Korean Astron.~Soc.}, {\bf 32}, 109.\par
\refn Svechnikov, M.A., Kuznetsova, E.F., 1990, {\it Katalog priblizhennykh fotometricheskikh i absoliutnykh
elementov zatmennykh peremennykh zvezd}, Izdatelstvo Ural'skogo universiteta, Sverd\-lovsk.\par
\refn Vijapurkar, J., Drilling, J.S., 1993, {\it Astrophys.~J.~Suppl.}, {\bf 89}, 293.\par
\refn Walker, G.A.H., {\it et al}. 2005, {\it Astrophys.~J.}, {\bf 623}, L145.\par
}
\end{document}